\documentclass[11pt,twoside]{article}
\usepackage{asp2010}
\usepackage{psfig,graphicx}

\resetcounters

\begin{document}

\title{YouASTRO: a web-based bibliography management system with distributed comments and rating features for SAO/NASA ADS papers}
\author{F. Bocchino$^1$, J. Lopez-Santiago$^2$, F. Albacete-Colombo$^3$ and N. Bucciantini$^4$
\affil{$^1$INAF-Osservatorio Astronomico di Palermo, Piazza del Parlamento 1, 90134 Palermo, Italy \\
$^2$Departamento de Astrofísica, Facultad de Ciencias Fisicas, Universidad Complutense de Madrid, E-28040 Madrid, Spain \\
$^3$Facultad de Ciencias Astronómicas y Geofisicas, Universidad Nacional de La Plata; Centro Universitario Regional Zona Atlantica (CURZA), Universidad Nacional del COMAHUE, Argentina \\
$^4$INAF-Osservatorio Astrofisico di Arcetri, Largo Enrico Fermi 5, Firenze, Italy}
}

\begin{abstract}
We present a working prototype of YouASTRO (www.youastro.org), a web-based BibTeX-compliant reference management 
software (RMS) for astrophysical papers in the SAO/NASA ADS database. 
It also includes as 
a main feature the concept of distributed paper comments and ratings. 
In these paper, we introduce the main characteristics of the web application, and we will briefly discuss what could be the advantages and drawbacks of 
such a system being widespread adopted by the astrophysical community 
for its scientific literature.
\end{abstract}

\vspace{-1cm}
\section{Introduction}
A reference management software (hereafter RMS), also known as citation management software or personal bibliographic management software, is a software which allows scientists to manage and organize list of references, especially for (but not limited to) their inclusion as citations in manuscripts written by them. Wikipedia lists 29 RMSs, of which 14 are under not-proprietary license\footnote{http://en.wikipedia.org/wiki/Comparison\_of\_reference\_management\_software}. Many of the software (18) are available only on few selected operating systems. From an astrophysicist point of view, among the most desired features of an RMS there are the support of BiBTeX both for import and export of the lists, and the possibility to easily include papers from the SAO/NASA Astrophysics Data System (\citealt{eag07}) and the ArXiv archive (\citealt{g11}), whereas other databases or formats are seldom used. A products free from license-fees would be more appealing than other solutions. YouASTRO (www.youastro.org) is a RMS which was developed by a group of astrophysicists to meet some of the needs that we encountered during our daily job. 

As an additional feature, we have implemented in YouASTRO the concept of distributed and shared comments. Participatory information sharing is now so widespread on Internet to be considered one of the key aspect of the so-called Web 2.0. In general, it is widespread recognized as one of the most attractive feature of the net, and a way to promote freedom of speech and dissemination of knowledge.

We stress that YouASTRO is not a replacement for ADS or astro-ph, it just sits on top of them. You cannot submit any paper to YouASTRO. The only papers who can be managed by YouASTRO are the publications present in the NASA/SAO ADS database\footnote{Since astro-ph arXiv papers are also inserted daily in the ADS database, they are can also be managed by YouASTRO.}.

In Sect. 2, we will lists some of the key features of YouASTRO and in Sect. 4 we will discuss some benefits and potential problems of the platform.

\section{YouASTRO key features}

\subsection{Managing list of papers}

\begin{figure}
  \centerline{\hbox{
     \psfig{file=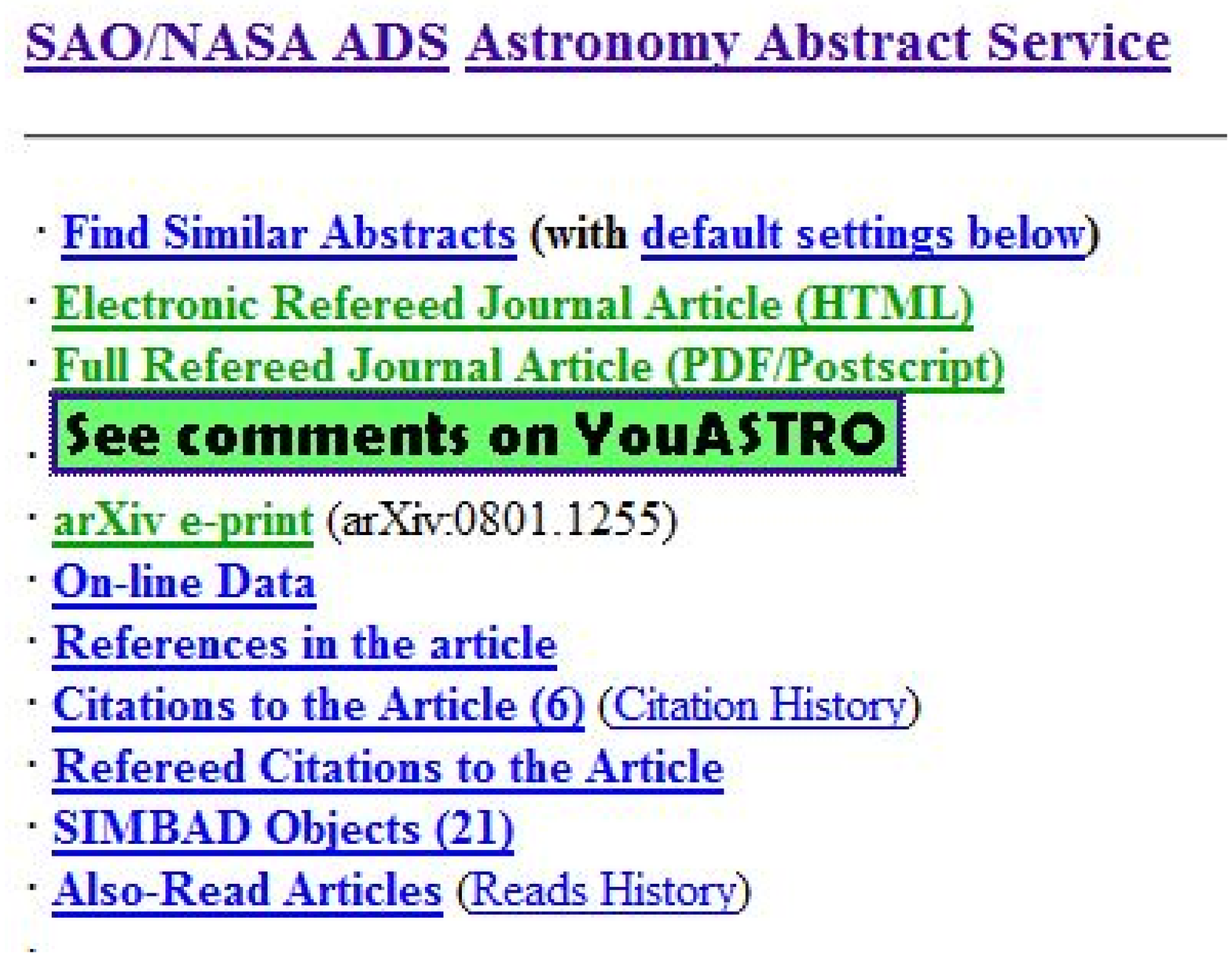,width=6.5cm}
     \psfig{file=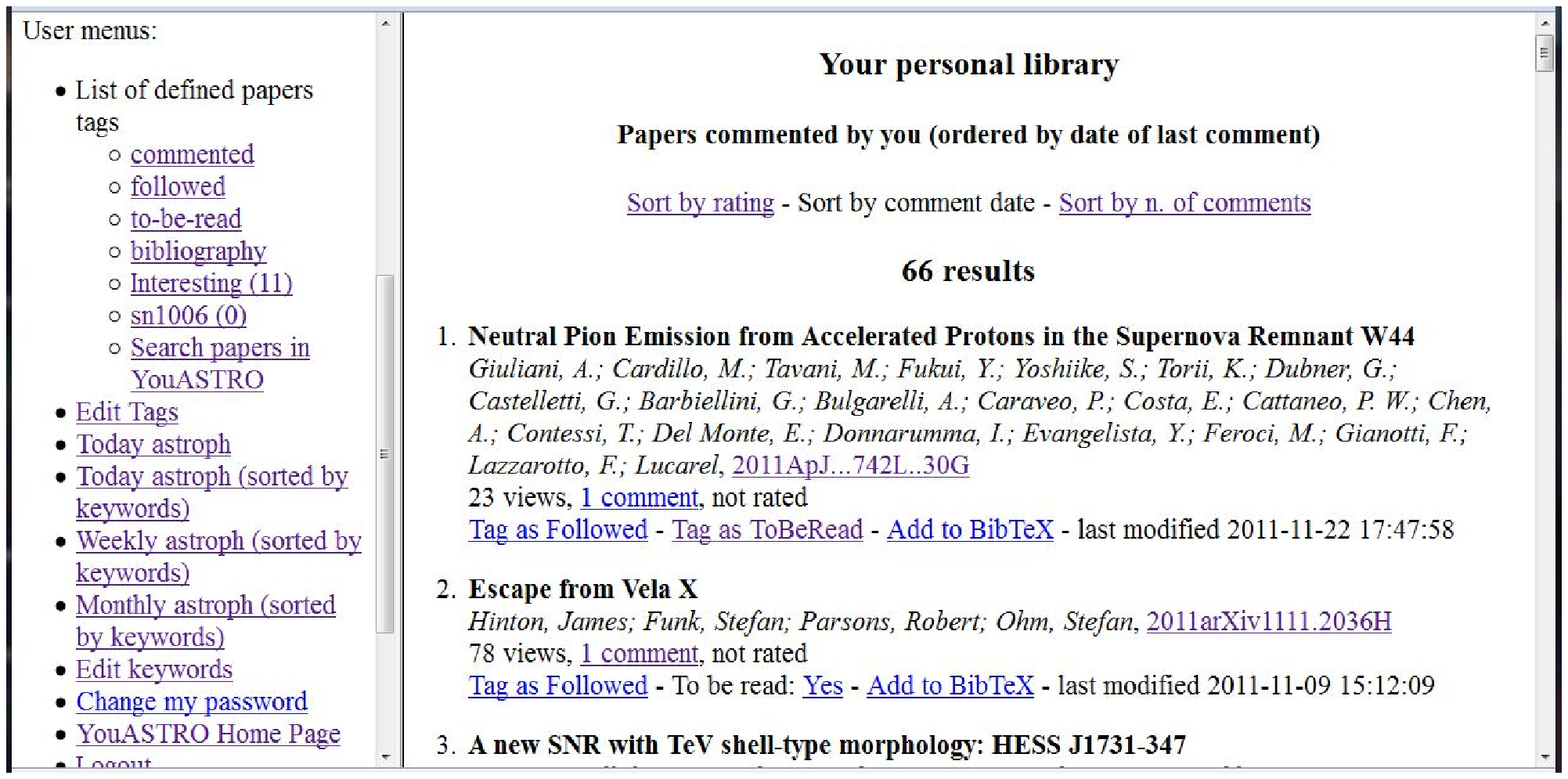,width=11.0cm}
  }}
  \caption{{\em Left Panel:} Screen dump showin the YouASTRO link appearing in ADS results, after we have set it up in the Libary Link section of myADS. {\em Right panel:} YouASTRO user home page. In the right frame: the list of tags, the edit tag link, and the links for getting the daily, weekly and monthly astro-ph digest filtered by keywords.}

  \label{ss1}
\end{figure}

YouASTRO has been designed to manage lists of papers by astrophysicist during their daily job. For now, the lists of papers which YouASTRO users can manage are the followings (see also Fig. \ref{ss1}, right panel):

\begin{itemize}
\checklistitemize
\item List of paper the user commented. This list is automatically updated whenever the user leave a comment to a paper (see next section about commentaries)
\item List of papers which a user want to cite in publications (a.k.a. "`my bibliography"'). This is a special list which can be exported as a single BiBTeX source file, for its use in a LaTeX manuscript. Papers in this list have an automatic bibliographic code generated when they are added. The code is composed by the surname first letter of the 3 first authors plus the final 2 digits of the publication years. Users can freely change the code of the papers in their bibliography.
\item List of papers tagged as "`Followed"'. If a paper is in this list, the user will receive a notification email when somebody else leave a comment on it. This is useful to keep an eyes to discussion and threads going on about specific papers. It is mostly useful, for example, to be informed about comments on a papers authored by the user.
\item List of papers tagged as "`To be read"'. This is a list of papers which we want to read sooner or later, but we haven't done it yet.
\item User-defined lists. This is similar to the concept of Personal Library of myADS. In YouASTRO the papers in the user-defined lists can be searched quickly using the "Search paper" form.
\end{itemize}

For each of this lists, users may add/remove papers with one click, search in the lists, sort according to various parameters (e.g. year of publication, number of comments, etc.), print, export in BibTeX, send to friends. YouASTRO allows you to do all these operations, just using your web browser, without the need of installing any additional software. 

Registered users may populate the lists quite easily in many ways, e.g. by activating YouASTRO in the Library Link feature of the NASA/SAO ADS. This would allow users to keep using the ADS query interface as usual and having a link to YouASTRO automagically appear in query results (Fig. \ref{ss1}, left panel). Other possibilities are browsing the integrated astro-ph keyword-filtered digests available in YouASTRO (a web-based alternative to the myADS keyword filtered mailings, Fig. \ref{ss1}, right panel), or doing an ADS search through the YouASTRO ADS search form.

%

\subsection{User commentaries and ratings}

YouASTRO gives the possibility to write commentaries to papers. The comments will appear on the YouASTRO record of a given publication. Moreover, the number of comments will also appear in the user lists of papers and on the results of a query done to ADS using the YouASTRO form developed for such purpose. Commentaries are not moderated, but a Board of Editors (BoE) will collect all the abuse report filed by users and take care of other Terms of Service violation (see Sect. 3).

Comments may be of three different kinds:

\begin{itemize}
\checklistitemize
\item Public and signed by the user. In this case, the comments may be read by anybody on the Internet, bu accessing the YouASTRO record for the paper. Anybody here means both YouASTRO registered users and not-registered users. The comment will be displayed with the corresponding YouASTRO username of the author of the comment. These kind of comments may be written, by definition, only by registered users, whose retain the rights to modify them and to delete them.
\item Public and anonymous. In this case, the comments may also be read by anybody, but the name of the author will not be neither displayed nor stored in the database. This means that the author of the comment will lose the right to modify or to delete the comment.
\item Private. Private comments are never displayed to anybody, but the comment's author, who can modify or delete them as he/she likes.
\end{itemize}

Registered user may also rate the paper on a scale from 1 to 10. Rate are always anonymous and the system will display the average rate and its standard deviations when more than 3 rates are available for a paper.

\section{Potential problems and benefits}

Scientist may be reluctant at first to accept a open-access system in which their work can be freely commented by anybody. In reality, these concerns are overestimated, because positive Internet experiences of participatory information sharing (e.g. Wikipedia) have shown that open-access communities tend to autoregolate quickly, and abuses are kept well under control.
In general, 3 types of abuses are envisaged: 1) page vandalism, 2) comments including insults, racism, gender discrimination, etc., and 3) not competent or not relevant comments. The types 1 and 2 are against the YouASTRO Term of Service (ToS) and they will be immediately removed upon user report and/or Boar of Editor (BoE) detection, and/or the triggering of an automated alert robot which have been setup for this purpose. IP addresses are always logged, so the BoE has the ability to ban an offending IP in case of repeated abuses. The third kind may be not against ToS, and they may be borderline in many cases. The BoE will act only after a user has submitted an abuse report using the link available in every comment, stressing that in many cases the most effective countermeasure of this kind of problem is just replying to the offending comment.

The potential benefits of the commentary system for the scientific research are enormous. Comments may be used to pinpoint strong and weak points of a paper,
to cope with cases of plagiarism (even borderline) and to remind authors about possible "forgotten" citations. These kind of problems have been traditionally faced with private emails and discussions, and having them pointed out in a commentary page will allow everybody to know the reasons of both parts and to help everybody to have an idea of what is going on. Comments are NOT intended to replace discussions with colleagues at meetings etc., NOR they are a replacement of the peer review system used by journals, but they are complementary to them.

YouASTRO promotes the online scientific discussions focused on papers, 
it is a way to leave on the web our points of view, for others to read them and to use them as they want, 
in the framework of a general and continuous improvement of the quality of scientific publications, and the overall advance of science.

The application is already available at the web site www.youastro.org. Comments and suggestions are welcome, email them to board@youastro.org.

\acknowledgements This paper was partially funded by the ASI-INAF contract I/009/10/0.

\bibliography{yaReferences}

\begin{thebibliography}{}
\expandafter\ifx\csname natexlab\endcsname\relax\def\natexlab#1{#1}\fi
\expandafter\ifx\csname url\endcsname\relax
  \def\url#1{\texttt{#1}}\fi
\expandafter\ifx\csname urlprefix\endcsname\relax\def\urlprefix{URL }\fi
\providecommand{\eprint}[2][]{\url{#2}}

\bibitem[{{Eichhorn} et~al.(2007){Eichhorn}, {Accomazzi}, {Grant}, {Henneken},
  {Kurtz}, {Bohlen}, {Thompson}, \& {Murray}}]{eag07}
{Eichhorn}, G., {Accomazzi}, A., {Grant}, C.~S., {Henneken}, E., {Kurtz},
  M.~J., {Bohlen}, E.~H., {Thompson}, D.~M., \& {Murray}, S.~S. 2007, in Lunar
  and Planetary Institute Science Conference Abstracts, vol.~38 of Lunar and
  Planetary Inst. Technical Report, 1240

\bibitem[{{Ginsparg}(2011)}]{g11}
{Ginsparg}, P. 2011, \nat, 476, 145

\end{thebibliography}

\end{document}